\documentstyle[preprint,aps,epsfig,eqsecnum]{revtex}
\draft 
\def\la{{\langle}} 
\def\ra{{\rangle}}  \def\y{\'\i} 
\newcommand{\beq}{\begin{equation}} 
\newcommand{\eeq}{\end{equation}} 
\newcommand{\beqa}{\begin{eqnarray}} 
\newcommand{\eeqa}{\end{eqnarray}} 
\newcommand{\da}{^\dagger}

\newcommand{\Om}{\Omega}

\newcommand{\I}{{\mathcal I}}
\newcommand{\G}{{\mathcal G}}

\def\half{{1\over 2}}

\def\non{\nonumber }
\def\beq{\begin{equation} }
\def\eeq{\end{equation} }
\def\beqa{\begin{eqnarray}}
\def\eeqa{\end{eqnarray}}
\def\del{\partial }
\def\a{\alpha }
\def\am{\alpha^\mu }
\def\Xm{X^\mu}
\def\Xn{X^\nu}
\def\d{\textrm{d}}
\def\b{\beta}
\def\tt{\tilde{\tau}}
\begin{document}

\title{Emission Spectrum of Fundamental Strings:\\ An Algebraic 
Approach} \author{J. L. Ma\~nes} 
\address{Departamento de F\y sica  de la Materia Condensada\\
 Universidad del Pa\'{\i}s Vasco,
 Apdo. 644, E-48080 Bilbao, Spain  \\
 {\bf wmpmapaj@lg.ehu.es}} 
\maketitle
\begin{abstract}
We  formulate  a linear difference equation which yields  averaged semi-inclusive 
decay rates for arbitrary, not necessarily large,  values of the masses. We show that 
the rates for decays $M\!\to\! m\!+\!M'$  of \emph{typical} heavy open strings are  independent 
of the masses  $M$ and $m$, and compute the 
 ``mass deffect'' $M\!-\!m\!-\!M'$. For closed  strings we find
decay  rates proportional to $M m_{R}^{(1-D)/2}$, where $m_{R}$ is the 
reduced mass of the decay products. Our method  yields exact  
interaction
rates valid for all mass ranges  and may  provide a fully microscopic 
basis, not limited to the long string approximation, for the interactions 
in the Boltzmann  equation approach to hot string gases.
  
\end{abstract}
\vskip1cm
\pacs{ \hfill EHU-FT/0105}
\newpage
\section{Introduction}

The decay 
properties of very massive strings were actively studied in the 
past~\cite{dec,nod,cdr,dyp} and several
general features, such as the existence of a constant  decay rate 
per unit length and the dependence   of the total decay on the mass of 
the string emerged. Many of those works extracted the decay 
rates from the imaginary part of the self-energy diagram and, for technical reasons, 
the computations    were   restricted to decays of very particular states, such as 
strings on the leading Regge trajectory~\cite{dec,nod,cdr} or strings 
stretched around a compact spatial dimension~\cite{dyp}. Moreover,  given 
that the one-loop diagram involves an
integral over the momenta of the decay products, no information was 
obtained on the energy distribution of the emitted strings. 

However, we can easily imagine situations where we don't know or are 
not interested in the full  details of the initial  
state. For instance, a high temperature string
gas will contain
strings of many different masses, and we may want to  find
 the corresponding  distribution function~\cite{stg,bar,bac}. 
One possible approach is to try to formulate and solve a  
Boltzmann equation for the system~\cite{hss,dbr}, and for 
that we need to know the decay (and recombination) rates of \emph{typical} 
strings of mass $M$. The averaged total interaction rate of two highly excited 
closed strings was computed in~\cite{lys} and, for massive long strings 
in different situations, one may usually guess the form  
of the rates from semiclassical considerations~\cite{hss,dbr}. But it 
would  be
interesting to have a fully  microscopic formulation valid for all 
mass ranges.

Similarly, a highly excited string is believed to form a black hole at 
strong coupling. But we can try to mimic this at weak coupling by 
pretending that we only know the mass of the state, and see if the 
properties obtained by taking averages over all the states of given 
mass $M$  resemble those of an actual black hole. In particular, we 
would like to compare  the energy and mass spectrum of the 
emitted strings.

\newpage

In this spirit, the photon emission  rate 
by (bosonic) massive open strings\footnote{They also studied massless  emission 
by  charged closed strings  
and obtained   grey body factors   of the type 
found in black holes~\cite{gbf}. We will consider closed strings in 
Section~4.}
 was considered in~\cite{ayr}. 
This can be 
formulated as an initial string state $|\Phi_{i}\ra$ going into a 
final state $|\Phi_{f}\ra$ through the emission of a photon   of momentum 
$k^\mu$ and,   
at weak coupling, the  amplitude for this process is proportional to 
the matrix element $\la\Phi_{f}|V(k)|\Phi_{i}\ra$ of the photon
vertex operator.\footnote{See, for 
instance~\cite{gsw,pol}.}  The resulting decay rate, which can
 be computed for any initial 
and final states, has in general a complicated polynomial 
dependence on $k^\mu$. However, the authors of~\cite{ayr}  showed 
that,  
when these  rates are \emph{averaged} over the many 
initial states $|\Phi_{i}\ra$ with mass $M$ and fixed momentum $p^\mu$ 
and \emph{summed} over all final 
states $|\Phi_{f}\ra$, the resulting spectrum is 
thermal. More precisely, they considered the averaged semi-inclusive emission rate
\beq\label{cuarentainueve}
d\Gamma_{N}(k^0)\propto {1\over {\cal G} (N)} \sum_{ \Phi_i|_{N}}\sum_{ \Phi_f|_{N'}} \big|
 \langle \Phi_f |  V(k)|\Phi_i \rangle \big|^2  k^{D-3} dk\ ,
\label{uno}
\eeq 
where the sums run over all states satisfying the mass conditions $\a'M^2= 
N-1$, \hbox{$\a'M'^2= N'-1$},
${\cal G} (N)$ is the degeneracy of the $N^{\mathrm th}$ mass level, 
and $D$ is the space-time dimension.
Evaluating (\ref{uno}) for large $N$ they obtained the following result for the 
photon emission rate
\beq\label{cuarentaisiete}
d\Gamma_N(k^0)\cong {\rm const.} \ 
{e^{-{k^0\over T_H}} \over 1-e^{- {k^0\over T_H}}  }\ k^{D-2} dk \ ,
\eeq
which is the spectrum of a black body at the Hagedorn temperature
\beq
T_{H}=\frac{1}{2\pi\sqrt{\a'}}\sqrt{\frac{6}{D-2}} \ .
\eeq

This identifies the Hagedorn temperature~\cite{hag} as the radiation 
temperature of an averaged string, and can be seen as an example of the 
emergence of classical or statistical concepts from microscopic 
computations through the average (decoherence) process~\cite{ayr,coh}.
It also shows that the properties of typical or averaged strings can be 
very different from those exhibited by specific states.

It is somewhat surprising that all  previous studies 
regarding the decays of fundamental strings missed this result, but
the calculation in~\cite{ayr} differs in two important points from 
those carried out in the past~\cite{dec,nod,cdr,dyp}. Namely, rather than 
fixing the initial state and summing over all final decay products, 
(\ref{cuarentainueve}) fixes the state of one of the emitted 
strings, and sums over all initial and final states of 
masses $M$ and $M'$. Also,  unlike  previous computations~(\ref{cuarentainueve})  does 
not sum over the momenta of the decay products, thus yielding   
the energy spectrum of the emitted photon. 
If we can evaluate~(\ref{cuarentainueve}) not only for photons, but 
for \emph{any} vertex operator $V(k)$, we will have complete 
information on the emisssion spectrum of typical strings. In 
particular we  will be 
able to compare the rates at which strings of different masses and 
spins (representations) are emitted. 

Having such a detailed 
information on the decay modes of typical strings can be useful in several ways. 
For instance, we may try too see 
which of  the features present in the decays of strings on the leading 
Regge trajectory~\cite{dec,nod,cdr,dyp}  apply also to typical strings. This 
is important, since states on the leading Regge trajectory are very 
atypical, maximum angular momentum states. 
Other possible applications include  a fully microscopical derivation of the interactions 
appearing in  
the Boltzmann equation approach to string gases~\cite{hss,dbr}, and a comparison 
between the decay properties of strings and black holes. For masses 
below the correspondence point~\cite{cor}, black holes transition to a 
random-walk-size string state or ``string ball''~\cite{dyv,rob}, which 
may even be accessible to production in the LHC or VLHC~\cite{rob}. If 
this turns out to be the case, computing its  decay properties is 
obviously relevant.  

These are the main motivations behind  this paper, where  we develop a 
method which allows the computation of 
the double sum in~(\ref{cuarentainueve}) for any  (not 
necessarily large) values of $N$ and $N'$, and for vertex operators 
corresponding to arbitrary physical states\footnote{For simplicity, 
only the bosonic string is considered in this paper.}. As we will see, a direct 
evaluation of~(\ref{cuarentainueve}) becomes very involved when 
$V(k)$ corresponds to a   massive string, but the problem can be 
reformulated in terms of a linear difference equation. This equation 
depends only on the mass of the emitted states, and different 
``initial conditions'' correspond to different states  
with the same mass. In this way we can neatly separate the universal 
properties of the decays from those dependent on the details of the 
emitted state. In particular, we can show that as long as the mass $m$ of 
the emitted string is less than $2M/3$ ($M$ is the mass of the 
decaying string), the emission rate is essentially independent of its 
state. The averaging process leaves only a 
universal dependence on~$m$. This is no longer true for $m\geq 2M/3$.

The solutions to the linear difference equation can be used to study
the decays of open and closed strings, both heavy and light. For 
example, we show 
that the rates for decays $M\to m+M'$  of heavy open strings 
are  independent of the masses  $M$ and $m$, and compute the 
mean value for the ``mass deffect'' $M\!-\!m\!-\!M'$. For closed  strings we find
decay  rates proportional to $M m_{R}^{(1-D)/2}$, where $m_{R}=m (M-m)/M$. Since 
our solutions are valid also for small values of $M$, we can see how the 
decay properties evolve as $M$ changes.

This  paper is organized as follows. 
In Section~2 we present a brief 
review of photon emission as computed in~\cite{ayr} and obtain a 
general formula which applies to the emission of arbitrary massive 
states. This formula involves  contour integrals over  a modified correlator 
on the cylinder, and a direct evaluation is difficult . However, in 
Section~3 we show that the integrals satisfy  a universal recursion 
relation, which depends only on the mass of the emitted particle, and 
compute and analyze the solutions. These solutions are used in 
Section~4 to  study the processes of fragmentation and radiation by heavy open and closed 
typical strings, and a comparison is made with previous results for 
strings on the leading Regge trajectory. We also analyze  decays of 
(relatively) light strings.
Our conclusions and outlook are presented in Section~5.

\section{Master Formula for emission  of arbitrary states }
\subsection{Review of photon emission}
The first step in the evaluation of the averaged 
inclusive rate in~\cite{ayr} is the transformation of  (\ref{uno}) into a double complex 
integral. To this end,  write the double sum as
\beq\label{treintaidos}
F(N,N')\equiv \sum_{ \Phi_i|_{N}}\sum_{ \Phi_f|_{N'}} \big|
 \langle \Phi_f |  V(k)|\Phi_i \rangle \big|^2= 
 \sum_{ \Phi_i|_{N}}\sum_{ \Phi_f|_{N'}} \langle \Phi_i |  V\da(k)|\Phi_f \rangle \langle 
 \Phi_f |  V(k)|\Phi_i  \rangle 
\eeq
and introduce projection operators over the mass levels
\beq\label{cincuentaiuna}
\hat P_N= \oint _C {dz\over z} z^ {\hat N-N}\ \ \ , \ \ \ \hat 
N=\sum_{n=1}^\infty \a_{-n}\cdot\a_{n}\ .
\eeq
With their help the sums can be converted to sums over all physical states, 
yielding 
\beq\label{dos}
F(N,N')= \oint_C {dz\over z} z^{-N  }  \oint _{C'} {dz'\over z'} 
{z'}^{-N ' }{\rm Tr}\ \big[z^{ \hat N } V\da(k,1) {z'}^{ \hat N } 
V(k,1)\big]\ ,
\eeq
where $C$ and $C'$ are small contours around the origin.
The vertex operator has been arbitrarily placed
 at $z=1$, 
although $SL(2,R)$ invariance of open string tree amplitudes 
guarantees that  the result is independent of 
this choice. For photons
\beq\label{quince}
V(k,z)=\textrm{i}\,\xi_\mu \del_\tau X^\mu (z) e^{ik.X(z)} ,\  \  \xi_\mu 
k^\mu=0 ,\   k^2=0\ ,
\eeq
where we have introduced the Euclidean proper time $\tau\equiv\ln z$ 
and $\Xm$ is given by the mode expansion\footnote{Henceforth we 
set $\alpha'=1/2$.} 
\beqa
\Xm(z)&=&x^\mu-ip^\mu\ln z+i\sum_{n=1}^\infty \frac{1}{n}\left(\am_{n} z^{-n}-\am_{-n} 
z^{n}\right)\equiv x^\mu-ip^\mu\ln z+X^\mu_{o}(z)\non\\  && 
[x^\mu,p^\nu]=i \eta^{\mu\nu}\ \ \ ,
 \ \ \ \ [\a_n^\mu , \a_{-m}^{\nu}]=
n\delta_{n+m}\eta^{\mu\nu}\ .
\eeqa
Notice that we have defined the oscillator part $X_{o}^\mu(z)$ for latter 
convenience.
The trace in~(\ref{dos}) is evaluated by coherent states techniques and, 
making the change of variables 
$v=z'$ and $w=zz'$, the result is\footnote{In order to obtain this 
result one should also introduce coherent states for the ghost variables and 
add the corresponding contribution to the number operator in the 
projectors~(\ref{cincuentaiuna}).}
\beq\label{cincuenta}
F(N,N')=\xi_{\mu}\xi_{\nu}\oint_{C_w} 
{dw\over w} w^{-N} f(w)^{2-D} \oint _{C_v} {dv\over v}  v^ {N-N'}
\left( p_\mu p_\nu+\eta_{\mu\nu} \Omega (v, w)\right)\ ,
\eeq
where
\beq
\Om(v,w)=\sum_{n=1}^\infty n \bigg( v^n +{w^n(v^n+v^{-n})\over 1-w^n} \bigg) 
\eeq
and $f(w)$ is related to the Dedeking $\eta$-function by
\beq
f(w)=\prod_{n=1}^\infty (1-w^n)=w^{-1/24} \eta\left(\frac{\ln w}{2 \pi 
i}\right)\ .
\eeq
Note also that the power of $f(w)$ in~(\ref{cincuenta}) reproduces the partition 
function for the open string, and its power expansion generates the 
mass levels degeneracies
\beq\label{veinticinco}
{\mathrm Tr}\ w^{\hat N}=\prod_{n=1}^\infty (1-w^n)^{2-D}=\sum_{N=0}^\infty 
{\cal G} (N) w^{N}\ .
\eeq 
The $v$-integral can be done trivially, and we are left with
\beq\label{cuatro}
F(N,N')=\xi ^2 \ (N-N') \oint 
 {dw\over w} {w^{-N'} f(w)^{2-D}  \over 1- w^{N-N'} }\  .
\eeq

For large $N'$ this integral can be computed by a saddle point 
approximation, with the main contribution coming from $w\sim 1$, 
where $f(w)$ may be approximated by 
\beq
f(w)\sim{\rm const. }\ 
(1-w)^{-1/2} e^{-{\pi ^2 \over 6 (1-w)} }\ .
\eeq
This, together with the asymptotic formula  for the degeneracy of mass level 
$N$
\beq\label{treintaitres}
{\cal G} (N)\sim \ N^{- {D+1\over 4}   } e^{a\sqrt{N} }\ ,\ \ 
\ \ \ a=2\pi\sqrt{D-2\over 6}\ ,
\eeq
leads to the black body spectrum~(\ref {cuarentaisiete})  mentioned in the Introduction.

The computation above considers only one of the  two cyclically inequivalent 
contributions to the three-point  
amplitude. However, these two contributions are related by world-sheet 
parity, and differ only by  a sign equal to the product of the 
world-sheet parities of the states. For the open bosonic string the 
parity of a state depends only on its mass level and is given by $(-)^{N}$. 
Thus for photon emission the relative sign between the two cyclically 
inequivalent contributions  is given by  $(-)^{N+N'+1}$ and, for $U(1)$ strings, the 
total amplitude is zero for $N+N'$ even. For other groups the decay 
amplitude for $N\to N_{1}+N_{2}$ should be multiplied by 
$\textrm{tr}[\lambda(\lambda_{1}\lambda_{2}+(-)^{(N+N_{1}+N_{2})}\lambda_{2}\lambda_{1})]$.
For the rest of the paper we will assume a $U(1)$ group, and keep in 
mind  that  decay rates vanish for $N+N_{1}+N_{2}$ odd.

Before considering the emission of arbitrary states, we would like to 
make a remark regarding the meaning of~(\ref{cuarentainueve}) for 
$D\neq 26$.
Since we are not considering a compactification, strictly speaking the theory  makes 
sense only for $D=26$. However, this is just a tree level computation, and 
it seems that as long as we restrict ourselves to the 
``transverse'', positive norm states generated by $D-2$ oscillators we should not 
find any inconsistency. This restriction is usually described as a 
``truncation'' of the theory, and was widely used in previous 
computations of decay rates~\cite{dec}.
At the critical dimension, the total number of propagating physical states~\footnote{By 
\emph{physical state} 
we mean one that satisfies the Virasoro conditions 
$(L_{n}-\delta_{n,0})|\Phi\ra=0,\  \forall n\geq 0$. Propagating states are physical 
states with positive norm.} 
coincides with those generated by the 24 transverse oscillators of 
the light-cone gauge. However, for $D<26$ there are some extra ``longitudinal''
propagating states which can not be obtained from $D-2$ oscillators, 
and~(\ref{veinticinco}) does not give the total degeneracy, only the 
number of ``transverse'' states\footnote{The problem of counting 
physical states for $D<26$ is expained in great detail in~\cite{poly}.}.  
This would not be a problem if the 
sums in (\ref{treintaidos}) counted only transverse states, as 
 the power of $f(w)$ in~(\ref{cincuenta}) would suggest.
However,  a explicit computation in Section~4 shows that at least some of the 
longitudinal  states are counted in $F(N,N')$. On the other hand, the recursion relation 
to be obtained in Section~3 applies only to 
the emission of transverse states. 
Thus,  it seeems that one can have an 
entirely consistent picture only at the critical dimension, even 
though only tree amplitudes are beeing computed. Although we will 
continue to write most of our formulas for arbitrary $D$, one should 
bear in mind that their validity is clear only for the critical 
dimension.

\subsection{Emission of arbitrary states}
As is well known, to any physical state
\beq\label{trece}
|\Phi\ra=\sum c_{\mu_{1}\ldots\mu_{m}}\a_{-n_{1}}^{\mu_{1}}\ldots\a_{-n_{m}}^{\mu_{m}}
|0;k\ra
\eeq
we can associate a vertex operator $V(k,z)$ by making the following 
substitutions~\cite{pol} 
\beq
\alpha_{-n}^\mu \rightarrow {\textrm{i}\ \partial_{\tau}^n X^\mu(z)\over (n-1)!}\ \ \ ,\ \ \
|0;k\ra\rightarrow  e^{ik.X(z)}\ 
\eeq
 and  normal ordering the result. The simplest such state is the 
tachyon 
\beq
V(k,z)= : e^{ik.X(z)}:\ , \ \ \  k^2=-m^2=2\ .
\eeq

Since (\ref{dos}) is valid for any physical state, provided that we 
substitute the  appropriate vertex operator, we can  
 use the same standard coherent states techniques for tachyon 
emission.  Instead of~(\ref{cincuenta}) we obtain
\beq\label{cinco}
F(N,N')=\oint 
{dw\over w} w^{-N} f(w)^{2-D} \oint  {dv\over v}  v^ {N-N'}\ 
\hat\psi(v,w)^{-2}
\eeq
where
\beq
\hat\psi(v,w)=(1-v)\prod_{n=1}^\infty{(1-w^nv)(1-w^n/v)\over 
(1-w^n)^2} \ .
\eeq

Obviously, the form of the function $\hat\psi(v,w)$ renders a direct evaluation of 
the $v$-integral rather non-trivial. This point will be addressed in 
the next Section. Here we will consider the problem of 
writing the equivalent of~(\ref{cinco}) for an arbitrary physical state 
in a systematic way. One key 
observation is that $\hat\psi(v,w)$ is related to the scalar correlator 
 on the cylinder
\beq\label{cuarentaicinco}
\la \Xm (1) \Xn (v) \ra=-\eta^{\mu\nu}\ln\psi(v,w)
\eeq
by 
\beq\label{cuarentaiseis}
\hat\psi(v,w)=\sqrt{v}\exp\left( {-\ln^2 v\over2 \ln w}\right)\ 
\psi(v,w)\ .
\label{seis}
\eeq
Similarly, we recognize that the integrand in~(\ref{cinco}) would correspond 
to a two-point amplitude on the cylinder if we made the substitution
\beq\label{siete}
f(w)^{2-D}\rightarrow f(w)^{2-D}\left(\frac{-2\pi}{\ln 
w}\right)^{D/2}\ .
\eeq

These are not accidents. In the operator formalism\footnote{See for instance 
chapter VIII of~\cite{gsw}.} the integrand for the two-point one-loop 
amplitude is given by
\beq
\int d^{D}p{\rm Tr}\ \big[z^{ L_{0} } V\da (k,1){z'}^{ L_{0} } 
V(k,1)\big]=f(w)^{2-D}\left(\frac{-2\pi}{\ln w}\right)^{D/2}\la V\da 
(k,1) \ V(k,v) \ra \ ,
\eeq
where $v=z'$ and $w=zz'$. The extra factors in~(\ref{seis}) 
and~(\ref{siete}) 
are the result of the momentum 
integral over the zero mode parts of the Virasoro generator $L_{0}$, 
which do not appear in the definition~(\ref {dos}) of $F(N,N')$. This implies  
that, for \emph{any} physical state, the evaluation of the trace in 
$F(N,N')$ will give
\beqa\label{ocho}
{\rm Tr}\ \big[z^{ \hat N } V\da (k,1){z'}^{ \hat N } V(k,1)\big]&=&{\rm Tr}\ \big[
{V'}\da (k,1) V'(k,v) w^{ \hat N }\big]\non\\ &=&f(w)^{2-D} \la {V'}\da (k,1) 
\ V'(k,v) \ra\ \  ,
\eeqa
where the prime over the vertex operators  means that the following 
substitution should be made
\beqa\label{nueve}
\del_{\tau}^n \Xm(v)&=&(x^\mu-ip^\mu\ln v)
\delta_{n,0}-ip^\mu\,\delta_{n,1}+\del_\tau^n \Xm_{o}(v)\non\\
&&\qquad  \rightarrow v^{\hat N}\del_{\tau}^n 
\Xm(1)v^{-\hat N}=x^\mu\,\delta_{n,0}-ip^\mu\,\delta_{n,1}+\del_\tau^n 
\Xm_{o}(v)\ .
\eeqa
The reason for this substitution is that, although $L_{0}$ is the generator of proper time 
translations
\beq
z^{ L_{0} }\, V (k,1) {z}^{ -L_{0}}=V(k,z)\ ,\ \ \ L_{0}=\frac{1}{2} 
p^2+\hat N\   ,
\eeq
the operator $\hat N$ does not affect the zero modes in $\del_{\tau}^n 
\Xm(1)$.
Note  that the prime in $V'(k,1)$ is irrelevant.

The  correlator in~(\ref{ocho}) can be computed by using 
\beq\label{diez}
\la \del_{\tau}^n \Xm_{o}(1) \del_{\tau}^m\Xn_{o}(v) \ra=-(-)^n\eta^{\mu\nu}
(v\del_{v})^{n+m}\ln\hat{\psi}(v,w)
\eeq
and the \emph{master formula} for emission of general states is simply 
given by
\beq
F(N,N')=\oint  {dw\over w} w^{-N} f(w)^{2-D} \oint  
{dv\over v}  v^ {N-N'}\ 
\la {V'}^{\dagger} (k,1) \ V'(k,v) \ra\ ,
\label{doce}
\eeq
which should be evaluated according to the  prescriptions 
in~(\ref{nueve}) and~(\ref{diez}). It is easy to check that the tachyon~(\ref{cinco}) 
and photon~(\ref{cincuenta}) sums  follow immediately from this formula. One should 
simply notice that 
\beq\label{dieciocho}
\la e^{-ik.X_{o}(1)} e^{ik.X_{o}(v)}\ra=e^{-k^2\ln\hat\psi(v,w)}=\hat\psi(v,w)^{m^2}
\eeq
and
\beq
\Om(v,w)=-(v\del_{v})^2 \ln \hat\psi(v,w)\ .
\eeq

We shall illustrate the use of the  prescriptions 
in~(\ref{nueve}) and~(\ref{diez}) with two further 
examples. The vertex operator for the symmetric rank-two tensor at mass
level $N=2$ is given by
\beqa\label{dieciseis}
V(k,z)&=&\xi_{\mu\nu}: \del_\tau X^\mu (z) \del_\tau X^\nu (z) e^{ik.X(z)}\!:\ \ ,\ \
 k^2=-m^2=-2\non\\ \xi_{\mu\nu}&=&\xi_{\nu\mu}\ ,\ \ \xi_{\mu}^{\ \ \mu}=  
 k^\mu\xi_{\mu\nu} =0\ ,\ \   \xi_{\mu\nu} \xi^{\mu\nu}=\frac{1}{2} \ ,
\eeqa
where the last condition on the polarization $\xi_{\mu\nu}$ guarantees 
that the state is normalized. We obtain
\beqa\label{treinta}
\la {V'}^{\dagger} (k,1) \ V'(k,v) \ra&=&\hat\psi(v,w)^2\non\\&\times& \left[( 
\xi_{\mu\nu}p^{\mu}p^{\nu})^2+4(\xi_{\mu\rho}\xi_{\nu}^{\ \ \rho}p^{\mu}p^{\nu})
\Om(v,w)+2\xi_{\mu\nu}\xi^{\mu\nu}\Om^2(v,w)\right] \ .
\eeqa
Similarly, for the rank-two antisymmetric tensor at mass level $N=3$
\beqa\label{diecisiete}
V(k,z)&=&\xi_{\mu\nu}: \del_\tau X^\mu (z) \del_\tau^2 X^\nu (z) e^{ik.X(z)}\!:\ \ ,\ \
 k^2=-m^2=-4\non\\ \xi_{\mu\nu}&=&-\xi_{\nu\mu}\ ,\ \  
 k^\mu\xi_{\mu\nu} =0\ ,\ \   \xi_{\mu\nu} \xi^{\mu\nu}=\frac{1}{2}
\eeqa
the result is
\beqa\label{cuarentaiocho}
\la {V'}^{\dagger} (k,1) \ V'(k,v) \ra&=&\hat\psi(v,w)^4\non\\&\times& \left[ 
\xi_{\mu\rho}\xi_{\nu}^{\ \ \rho}p^{\mu}p^{\nu}
{\ddot\Om}(v,w)+\xi_{\mu\nu}\xi^{\mu\nu}\big(\Om(v,w){\ddot\Om}(v,w)-{\dot\Om}(v,w)^2
\big)\right] \ ,
\eeqa
where the dots stand for derivatives with respect to proper time 
($\del_{\tau}=v\del_{v}$). Obviously a direct evaluation of the 
$v$-integral for these correlators would be very difficult.

We close this section by pointing out that more complicated vertex 
operators may require  evaluations of self-contractions between 
operators at coincident points. These are easily computed by using
\beqa\label{once}
\la \del_{\tau}^n \Xm_{o}(1) \del_{\tau}^m\Xn_{o}(1) \ra&=&\la \del_{\tau}^n \Xm_{o}(v) 
\del_{\tau}^m\Xn_{o}(v) \ra\non\\ &&\;\;\;\;\;\; =-(-)^n\eta^{\mu\nu}(v\del_{v})^{n+m}
\ln\hat{\psi}_{R}(v,w)\Big|_{v=1} \ ,
\eeqa
where 
\beq
\hat{\psi}_{R}(v,w)=\frac{\hat{\psi}(v,w)}{(1-v)} \ .
\eeq
Note that the denominator  cancels the zero  at $v=1$ and all the 
derivatives appearing in~(\ref{once}) are non-singular. This simply reflects 
the fact that vertex operators are normal ordered.

\section{Recursion Relation}
\subsection{The derivation}
In this section we address the problem of computing the $v$-integral 
in the master formula~(\ref{doce}). We will show  that, if we rewrite the 
master formula in terms of  new functions $\I_{n}(w)$ 
\beqa
\I_{n}(w)&\equiv&\oint_{C_{v}}  {dv\over v}  v^n \la {V'}\da (k,1) \ 
V'(k,v) \ra\non\\F(N,N')&=&\oint _{C_{w}} 
{dw\over w} w^{-N} f(w)^{2-D} \I_{ N-N'}(w)\ ,
\label{diecinueve}
\eeqa
then the properties of the correlator    imply a  of recursion relation 
for  $\I_{n}(w)$. 
In order to simplify the derivation, we will write the vertex 
operators in a very specific gauge. Namely, the coefficients $c_{\mu_{1}\ldots\mu_{m}}$
in~(\ref{trece}) will be orthogonal to the momentum $k^\mu$
\beq\label{catorce}
k^{\mu_{1}}c_{\mu_{1}\mu_{2}\ldots\mu_{m}}=k^{\mu_{2}}c_{\mu_{1}\mu_{2}\ldots\mu_{m}}=
\cdots= k^{\mu_{m}}c_{\mu_{1}\mu_{2}\ldots\mu_{m}}=0 \ .
\eeq
Note that the vertex operators~(\ref{quince}), (\ref{dieciseis}) 
and~(\ref{diecisiete}) are written in this gauge. 
Although for photons~(\ref{catorce}) is a consequence of the physical state 
conditions, it becomes a gauge choice for  states 
with $N\geq 2$. This  gauge was analyzed in detail in~\cite{vop}, 
where it was shown that for each propagating (i.e.  
positive norm) physical state there is a 
\emph{unique} gauge representative satisfying~(\ref{catorce}) for 
$D=26$. For $D<26$ there are positive norm, physical ``longitudinal'' states which 
can not be written in this gauge, and our formalism is not useful in 
computing their emission rates.

For vertex operators satisfying~(\ref{catorce}) the only contraction involving 
the exponents is given by eq.~(\ref{dieciocho}). All other contractions will involve at 
least two proper time derivatives, and the correlator in $\I_{n}(w)$ will take the following 
form
\beq\label{veinte}
\la {V'}^{\dagger} (k,1) \ V'(k,v) 
\ra=\hat\psi^{m^2}{\mathcal P}(\Om,\dot{\Om},\ddot{\Om},\ldots) \ ,
\eeq
where ${\mathcal P}$ is a polynomial (cf.~(\ref{cincuenta}), (\ref{cinco}), 
(\ref{treinta}) and (\ref{cuarentaiocho})).

On the other hand, the contours in~(\ref{diecinueve}) must satisfy $|w|<|v|<1$, which 
follows from $|z|,|z'|<1$ and $v=z'$, $w=zz'$. The functions  $\I_{n}(w)$ 
can be viewed as the coefficients in the Laurent expansion of the correlator
\beq
\la {V'}^{\dagger} (k,1) \ V'(k,v) \ra=\sum_{n=-\infty}^\infty 
\I_{n}(w)\, v^n
\eeq
and this expansion is valid only inside 
an annulus between the nearest singularities, which 
are the poles at $v=1$ and $v=w$.\footnote{A general account of this 
and other properties of the correlator, as well as its relation to 
Jacobi theta functions can be found in  Appendix~A.}

\begin{figure}[!t]
\epsfysize=8cm
\centerline{\epsfbox{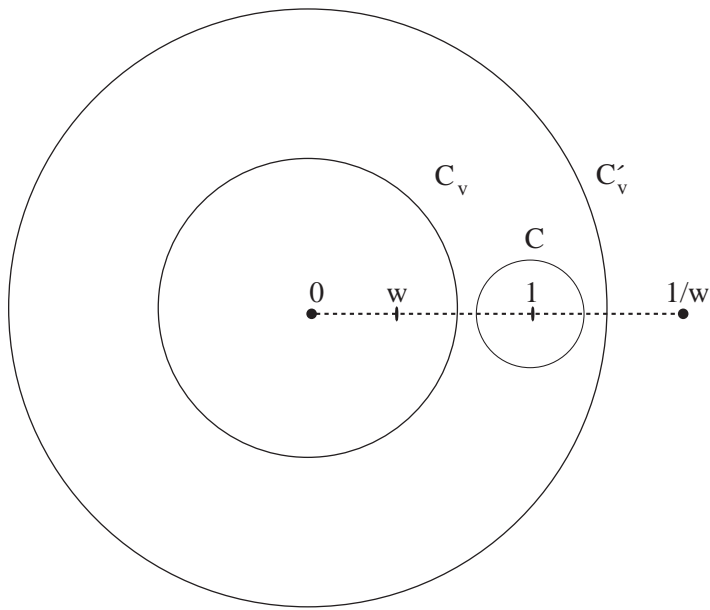}}
\caption{Contours used in the derivation of the recursion 
relation(\ref{veintidos}).}
\end{figure}

Let us consider  a second $v$-contour  $C_{v}'$ defined by 
$1<|v'|<|w|^{-1}$, which is related to $C_{v}$ by the transformation
$v'=w^{-1} v$. Using the  properties
\beqa
\hat\psi(w^{-1}v,w)&=&-\frac{v}{w}\hat\psi(v,w)\non\\
\del_{\tau}^{n}\Om(w^{-1}v,w)&=&\del_{\tau}^{n}\Om(v,w)
\eeqa
which follow directly from the definitions of $\hat\psi(v,w)$ and 
$\Om(v,w)$ and taking into account the general form of the correlator~(\ref{veinte}), we 
may relate the two contour integrals
\beqa
\oint_{C_{v}'}  {dv\over v}  v^n \la {V'}\da (k,1) \ V'(k,v) 
 \ra&=&w^{-n-m^2}\oint_{C_{v}}  {dv\over v}v^{n+m^2}\la {V'}\da (k,1) \ 
 V'(k,v)\ra\non\\ 
 &&\ \ \ \ \ \ \ =w^{-n-m^2}\I_{n+m^2}(w) \ .
\eeqa

The difference between the two contours can be deformed into a small 
contour $C$ around $v=1$  (see Fig.~1), and we have
\beqa\label{veintiuno}
\left(\oint_{C_{v}'}-\oint_{C_{v}}\right){dv\over v}  v^n \la {V'}\da (k,1) \ V'(k,v) 
 \ra&=&w^{-n-m^2}
\I_{n+m^2}(w)-\I_{n}(w)\non\\ &&\ \ \ =\oint_{C}  {dv\over v}  
v^n \la {V'}\da (k,1) \ V'(k,v) \ra \ .
\eeqa
This last integral depends only on the structure of the singularity 
at $v=1$, and is computed  in Appendix~A. Assuming that the emitted 
state is normalized, the result is
\beq\label{cuarentaicuatro}
\oint_{C}  {dv\over v}  v^n \la {V'}\da (k,1) \ V'(k,v) 
\ra=n+\frac{m^2}{2}
\eeq
and, after multiplying~(\ref{veintiuno}) by $w^{n+m^2}$, the following recursion 
relation is obtained
\beq\label{veintidos}
\I_{n+m^2}(w)=w^{n+m^2}\big[\I_{n}(w)+n+\frac{m^2}{2}\big] \ .
\eeq

This relation is \emph{universal}, in the sense that it depends only on the 
mass of the emitted state. By a similar argument involving two contours related 
by the transformation $v'=w/v$  the following ``reflection symmetry''  is 
easily derived
\beq\label{veinticuatro}
\I_{n}(w)=w^n\I_{-n}(w) \ .
\eeq

\subsection{The solutions}
For photons, the recursion relation reduces to 
$\I_{n}(w)=w^{n}\big[\I_{n}(w)+n\big]$ with the obvious solution
\beq
\I_{n}(w)=\frac{n w^n}{1-w^n} \ ,
\eeq
which agrees with~(\ref{cuatro}) and gives rise to the Plank distribution. For 
$m\neq0$ there are no closed-form solutions, but they can be expressed 
as  power power series in $w$. Since $\hat\psi(v,w)$ and all proper 
time derivatives of $\Om(v,w)$ 
have finite limits for $w\rightarrow0$, only \emph{positive} powers 
of~$w$ are involved. The absence of negative powers of $w$ combined 
with  the reflection symmetry~(\ref {veinticuatro}) 
implies 
\beq\label{cincuentaicuatro}
\I_{n}(w)=a_{n}w^n+O(w^{n+1})
\eeq
One can easily check that, for tachyons 
($m^2=-2)$, this condition implies that  
the \emph{unique} solution to~(\ref{veintidos}) is given by the series 
\beq
\I_{n}(w)=\sum_{p=1}^\infty (n-1+2p)w^{(n-1)p+p^2}\ .
\eeq

For positive $m^2=2,4,6,\ldots$ we can  use~(\ref{veintidos}) to express 
$\I_{n}(w)$ in terms of $\I_{n-m^2}(w)$, which  in turn can be related
to $\I_{n-2 m^2}(w)$ and so on, until one reaches $\I_{\nu}(w)$, where
$0\leq\nu<m^2$. The result of this iterative process is the following series
\beq\label{veintitres}
\I_{n}(w)=\sum_{p=1}^A \Big(n-m^2(p-\half)\Big)w^{np-\half 
m^2(p^2-p)}+\I_{\nu}(w)\,w^{\nu A+\half m^2(A^2+A)}\ ,
\eeq
where $\nu\equiv n \bmod{ m^2}$ and $n=m^2 A+\nu$. At first sight the 
solution~(\ref{veintitres}) depends on the $m^2$ undetermined 
functions $\I_{0},\ \I_{1},\cdots \I_{m^2-1}$. However, combining the 
reflection symmetry~(\ref{veinticuatro}) with the recursion relation yields
\beq
\I_{\nu}(w)=w^{2\nu-m^2}\I_{m^2-\nu}(w)+\big(\nu-\frac{m^2}{2}\big)\,
w^{\nu}\ ,
\eeq
which implies that only $\I_{0},\ \I_{1},\cdots \I_{m^2/2}$ are 
independent functions.

Note that the structure of this solution is consistent with the fact 
that~(\ref{veintidos}) is an inhomogeneous linear  difference equation. The last piece 
in~(\ref{veintitres}) is the general solution to the homogeneous equation
$\I_{n+m^2}(w)=w^{n+m^2}\I_{n}(w)$, which exists only for $m^2>0$; the 
sum from $p=1$ to $A$ is a particular solution to the complete equation.

As mentioned above, the recursion relation is universal, in 
the sense that only the mass of the emitted particle enters it. 
However, the solutions depend on the  functions $\I_{0},\ 
\I_{1},\cdots \I_{m^2/2}$, which should be determined  by direct 
expansion of the correlator. Since this expansion depends on the 
particular vertex operator considered, we may say that \emph{the solution 
to the homogeneous equation contains the non-universal, state dependent 
part of the emission rate, while the particular solution reflects 
the universal behavior.} The non-universal terms exist only for 
$m^2>0$. This is physically sensible, since there is only one 
tachyonic state and all $D-2$ transverse physical photon polarizations 
are equivalent by $SO(D-2)$ invariance.

In order to compare  the relative strength of universal and 
non-universal contributions to the decay rate, we should note 
that if  $\I_{n}$ is given by its power expansion 
\hbox{$\I_{n}(w)=\sum_{l>0}a_{l}\,w^l$},  
then by~(\ref{veinticinco}) and~(\ref{diecinueve}) $F(N,N')$ will be given by
\beq
F(N,N')=\sum_{l>0}a_{l}\;{\cal G} (N\!-\!l)\ .
\eeq
Since the degeneracy ${\cal G} (N)$ is a rapidly growing function of $N$,
the higher powers of $w$ will be strongly suppressed. On the other 
hand, the highest power in the universal series in~(\ref{veintitres}) is 
the term with $p=A$, which gives $l=\nu A+\half m^2(A^2+A)$. This is precisely 
the power of $w$ which multiplies $\I_{\nu}(w)$  in the non-universal term and 
since  $\I_{\nu}(w)=O(w^\nu)$,  
we see that the non-universal term 
will tend to be suppressed in relation to the leading universal contributions. 
This argument fails only for $A=0$ 
which, as we will see in the next Section, corresponds to the 
emission of  states carrying a large fraction of the total mass $M$.

\section{Applications}

In this Section we will describe some applications of the solutions 
to the recursion relations.  Substitution of the solution~(\ref{veintitres}) 
into~(\ref{diecinueve}) gives the  sum over initial and final states
as a linear combination of level degeneracies
\beq\label{veintisiete}
F(N,N')=\sum_{p=1}^A 
\Big(n-m^2(p-\half)\Big)\G\Big[N-np+\half 
m^2(p^2-p)\Big]+F_{NU}(N,N')\ ,
\eeq
where $n=N-N'$ and $F_{NU}$ is the contribution from the non-universal term
\beq\label{cincuentaitres}
F_{NU}(N,N')=\oint _{C_{w}} 
{dw\over w} w^{-N} f(w)^{2-D}\I_{\nu}(w)\,w^{\nu A+\half m^2(A^2+A)}\ .
\eeq
Note that~(\ref{veintisiete}) is also valid for photons and tachyons if we 
set $A=\infty$ and $F_{NU}=0$. The mass level degeneracies can be 
obtained by direct expansion of the partition 
function~(\ref{veinticinco}) or from an approximate formula, given in 
Appendix~B, which is more 
accurate than the usual asymptotic expression~(\ref{treintaitres}).

Taking into account the normalization of the initial state and the two-body 
phase space, the \emph{averaged semi-inclusive} decay rate is given by
\beq\label{treintaicinco}
\Gamma(M\to m+M')=A_{D}\frac{g_{o}^2}{M^2}{F(N,N')\over {\cal G} 
(N)}\,k^{D-3}\ ,
\eeq
where
\beq
A_{D}=\frac{2^{-D}\pi^{\frac{3-D}{2}}}{\Gamma(\frac{D-1}{2})}
\eeq
and  $g_{o}$ is the open string coupling constant. Remember that the 
label ``averaged semi-inclusive'' refers to the fact that~(\ref 
{treintaicinco}) describes the production of strings of mass~$m$ in a 
very \emph{specific} state, by  \emph{typical} strings of mass~$M$,  
while summing over all final states of mass~$M'$.

\subsection{Fragmentation of heavy open strings}
Here we will assume that the mass of the initial string is so large 
that we can use the following asymptotic approximation for the  mass
level degeneracies
\beq\label{treintaicuatro}
\G(M)\sim(2\pi 
T_{H})^{-\frac{D-1}{2}}M^{-\frac{D+1}{2}}e^{M/T_{H}}\ \ ,\ \ \
T_{H}=\frac{1}{\pi}\,\sqrt{\frac{3}{D-2}}\ .
\eeq
This is~(\ref{treintaitres}) written in terms of the mass, 
and we have specified the value of the multiplicative constant  in order to get quantitative 
results. Using~(\ref{treintaicuatro}) and
\beq\label{cuarentaidos}
n=N-N'=M\,k^0-\half m^2\geq m(M-\half m)\ ,
\eeq  
which follows from $p=p'+k$ and the mass shell conditions, we can obtain the following estimate 
for the ratio between the first 
two terms in~(\ref{veintisiete})
\beq
\frac{\G(N-2n+m^2)}{\G(N-n)}\sim\exp\Bigg[-\frac{m}{T_{H}}\left(\frac{M-3m/2}{M-m}\right)\Bigg]\ .
\eeq
Obviously, as long as $0<m<2M/3$   the 
contribution of all the terms but the first can be neglected, and we 
may take
\beq\label{treintaiseis}
F(N,N')\approx (N-N'-\half m^2)\ \G(N')\ .
\eeq

If we further asssume that  $m\ll M$ we can use the approximation $n=N-N'\approx Mk^0$ 
together with 
\hbox{$\G(N')/\G(N)\approx e^{-k^0/T_{H}}$}. 
Then~(\ref {treintaicinco}) implies the energy distribution
\beq\label{treintaisiete}
d\Gamma(k^0)\approx A_{D}\,g_{o}^2\,e^{-k^0/T_{H}}k^{D-2}d k\ ,
\eeq
where we have used $dn=Mdk^0$ and $k^0dk^0=kdk$. This confirms that 
massive strings with $m\ll M$ are radiated according to a 
Maxwell-Boltzmann distribution which is totally \emph{independent} 
of the mass $M$ of the decaying string. 
Had we included the rest of the terms 
in the series in $F(N,N')$ we would have obtained Plank 
distribution. But even for the lowest lying massive states 
$k^{0}\geq m$ is large compared with $T_{H}$, and the difference between 
the two distributions can be neglected.

In what follows we are going to lift the restriction $m\ll M$ and 
consider instead the splitting of a string of mass $M$ into arbitrary 
states of masses $m$ and $M'$.
Note that~(\ref{treintaicinco}) gives the average rate of decay of a string 
of mass $M$ into a \emph{specific} state of mass $m$. However, our 
approximation~(\ref{treintaiseis}) for $F(N,N')$ implies  that all the states of a 
given mass level $m$ are emitted at the 
same rate. Thus, if we multiply the rate by $\G(m)$ the result will 
be the \emph{total}  rate for decays into arbitrary states with 
masses $m$ and $M'$
\beq\label{cuarenta}
\Gamma_{tot}(M\to m+M')\approx 
A_{D}\frac{g_{o}^2}{M^2}(N-N'-\frac{m^2}{2})\frac{\G(m)\G(M')}{\G(M)}\,k^{D-3}\ .
\eeq
Note that, unlike~(\ref{treintaiseis}), this formula is valid even 
for $m>2M/3$. The reason is that~(\ref{cuarenta}) is symmetric under 
interchange of the emitted strings, and at least one of the two 
emitted masses is going to be less than $2M/3$. The only conditions for the validity 
of~(\ref{cuarenta}) are that  $M$ sould be large and neither $m$ nor $M'$ should be zero.

As long as $m\gg 1$ and $M'\gg 1$, we can use~(\ref{treintaicuatro}) 
to write
\beq
\frac{\G(m)\G(M')}{\G(M)}\sim (2\pi T_{H})^{-\frac{D-1}{2}}\Bigg(\frac{m M'}
{M}\Bigg)^{-\frac{D+1}{2}}e^{-t/T_{H}}\ ,
\eeq
where $t\equiv M\!-\!m\!-\!M'$ is the total kinetic energy or ``mass 
deffect'' for the decay. Since the exponential dependence on $t$ 
implies $t\sim O(T_{H})\ll M$, we can also make the 
following approximations
\beq
N-N'-\frac{m^2}{2}\approx m M'\ \ \ ,\ \ t\approx\half\Big(\frac{M}{m 
M'}\Big)\,k^2
\eeq
and find for the total decay rate 
\beq\label{treintaiocho}
\Gamma_{tot}(M\to m+M')\approx A_{D}g_{o}^2\frac{(\pi 
T_{H})^{-\frac{D-1}{2}} 
}{2mM'}t^{\frac{D-3}{2}}e^{-t/T_{H}}	\ .
\eeq

Thus we obtain a Maxwell-Boltzmann distribution for the \emph{total} 
kinetic energy. Note that if we only measure the energy of \emph{one} of the 
decay products (say the one with mass $m$) we will find a temperature
\beq
T_{m}=T_{H}\bigg(1-\frac{m}{M}\bigg)\ .
\eeq
Of course, this is simply a ``recoil effect'', and in the limit $m\ll M$ we recover 
our previous result~(\ref{treintaisiete}). Another consequence of the 
distribution~(\ref{treintaiocho}) is that the average kinetic energy 
released in a decay is  given by 
\beq\label{treintainueve}
< t>=\half (D-1)T_{H}\ .
\eeq
In other words, as long as neither of the final masses is too small, 
the energy released in the decay $M\to m+M'$ is 
independent of the masses and satisfies the 
equipartition principle in $D-1$ spatial dimensions.

Another quantity that we may compute from (\ref{treintaiocho}) is the 
total rate of production of strings of mass $m$. We simply have 
to sum over all values of $N'$ 
using $dN'=-M'dt$, and we get\footnote{Since a bosonic string can decay into 
 \emph{heavier} states throught tachyon emission, the total rate 
diverges. Following~\cite{dec,cdr} we  avoid the problem by 
imposing $N'\leq N$ by hand, i.e. by restricting ourselves to 
non-tachyonic decays.} 
\beq
\Gamma_{m}=\sum_{N'} \Gamma_{tot}(M\to m+M')\approx A_{D}g_{o}^2\frac{(\pi 
T_{H})^{-\frac{D-1}{2}} }{2m}\int_{0}^\infty \d t t^{\frac{D-3}{2}}e^{-t/T_{H}}
=\frac{A'_{D}g_{o}^2}{m}\ ,
\eeq
where $A'_{D}=\pi^{2-D} 2^{-(D+1)} $. In terms of the level density $\rho(m)\!=\!m$, which 
follows from the mass shell condition, there are $\rho(m)\,dm$ levels 
with masses between $m$ and $m+dm$. Thus the number of particles 
produced within that interval is given by the density
\beq\label{cuarentaiuna}
\frac{\d\Gamma(m)}{\d m}=\Gamma_{m}\,\rho(m)\approx A'_{D}g_{o}^2 \ ,
\eeq
which is independent of the masses. This means  that a heavy open 
string radiates strings of all masses with equal probabilities  and  implies a 
total decay rate proportional to the mass of the string. Thus, the decay of a heavy open string 
can be  described as a process of fragmentation, rather than   one of 
radiation.

From a semiclassical point of view, this can be explained in terms  of a 
constant splitting probability per unit of length~\cite{cdr,dyp}. Since the released 
energy~(\ref{treintainueve}) is small compared to the masses, $M\approx m+M'$ and all 
masses smaller than 
$M$ are produced with the same probability. This should fail when one 
of the masses is comparable to the released energy. From the point 
of view of our computation, this failure can be attributed to the breakdown of the 
asymptotic approximation for the level multiplicity.

The total decay rate of a string of mass $M$ is obtained by 
integrating (\ref{cuarentaiuna}) from $0$ to $M/2$, to avoid counting 
each decay mode twice. The result for $D=26$ is $g_{o}^2 \pi^{-24}2^{-28} M$
and can be compared to other computations in the literature. The 
decay rate for open bosonic strings on the leading trajectory is 
given by eq.~(12) of~\cite{cdr} and exactly agrees 
with ours. They obtain their result from the 
one-loop \emph{planar} diagram, which is equivalent to considering 
only one cyclic ordering for the three-point tree amplitude. Including 
the non-planar contribution as in~\cite{ayr} amounts  to 
adding the other cyclic ordering, but the  result is easy to 
predict. As mentioned in Section~2,  amplitudes for $U(1)$ 
strings  vanish or get a factor of~$2$ depending on the parity of $N+N_{1}+N_{2}$. 
The net result is 
that the total decay rate should  be multiplied 
by~$2$, yielding $g_{o}^2 \pi^{-24}2^{-27} M$.
This can also be compared with the decay rate for a long straight 
open string\footnote{Our oriented $U(1)$ string should be equivalent 
to the unoriented $SO(2)$ string considered in~\cite{dyp}. } , 
which according to eq.~(14) of~\cite{dyp} is given by 
$g_{o}^2 \pi^{-24}2^{-26} M$. This is twice our result, but the 
discrepancy may be due to some overlooked subtlety. In any case, 
since we are computing the \emph{averaged} decay rate for strings of 
mass $M$, the result need not coincide with those obtained for highly 
atypical states such as  those on the leading Regge trajectory.

We can also use~(\ref{cuarenta}) to compute the production rate $\Gamma_{m}$ by strings 
of mass $M$  numerically and compare the results with the asymptotic 
prediction~(\ref{cuarentaiuna}). Fig.~2 shows our results for $N=20$ and $N=100$ 
in $26$ dimensions, with $\frac{d\Gamma(m)}{d m}$ normalized so that $1$ 
corresponds to the asymptotic value~(\ref{cuarentaiuna}).
Note that even for $N$ as low as $20$ the rate manages to attain the predicted 
value for $m\approx 2.3$ and that for $N=100$ there is already a well 
defined plateau, with a decrease in the rate near the ends of the 
mass range.

\begin{figure}[!t]
\epsfysize=4.5cm
\centerline{\epsfbox{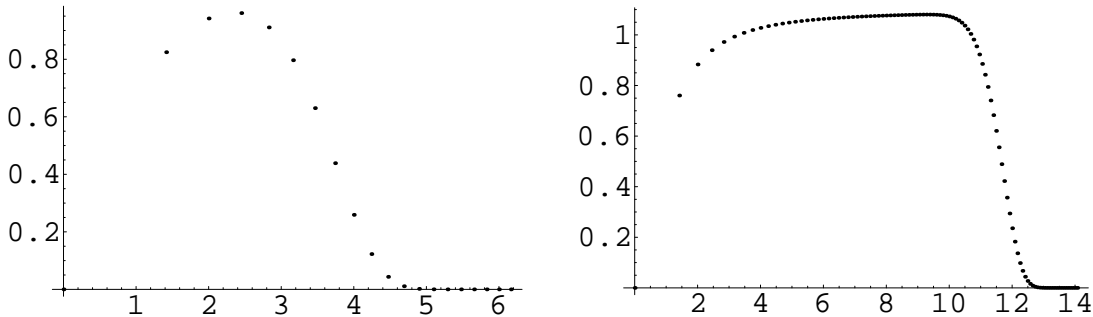}}
\caption{$\frac{d\Gamma(m)}{d m}$ for $D=26$,  $N=20$ ($M\approx 6.16$, left) and 
$N=100$ ($M\approx 14.1$, right).}
\end{figure}

For $D\neq 26$ the results do not match~(\ref{cuarentaiuna}) so well. For $D<26$ the production 
of light strings is  enhanced, while the opposite is true for $D>26$. 
This is illustrated in Fig.~3, which suggests that  strings 
decay
according to a the constant splitting probability only in $26$ 
dimensions~\cite{cdr}. 
However, one should not forget that the consistency of the formalism is suspect for 
$D\neq 26$. Since $\G(N)$ does not count the total 
number of propagating states for $D<26$, and for $D>26$ there are 
also negative norm physical states, the validity of 
(\ref{cuarenta}) for $D\neq 26$ is unclear.

\begin{figure}[h]
\epsfysize=3cm
\centerline{\epsfbox{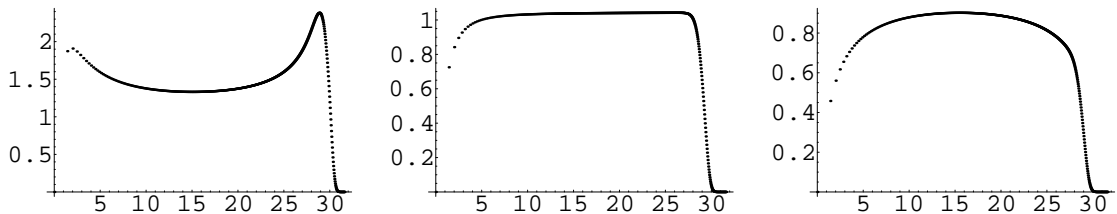}}
\caption{$\frac{d\Gamma(m)}{d m}$  for $N=500$  and $D=16$ (left),  $D=26$ (center) 
and $D=30$ (right). }
\end{figure}

\subsection{Radiation by heavy closed strings}
Our previous computations extend easily to the case of  heavy closed 
strings but, as we shall see, the results are remarkably different. 
Tree amplitudes for closed strings factorize into a product of left 
and right open string amplitudes , and instead of~(\ref{treintaicinco}) we have 
\beq
\Gamma_{c}(M\to m+M')=A_{D}\frac{g_{c}^2}{M^2}{F(N,N')^2\over {\cal 
G} (N)^2}\,k^{D-3}\ ,
\eeq
where $g_{c}$ is the closed string coupling constant, which is 
proportional to $g_{o}^2$. For closed strings, the mass shell condition becomes 
$\a'M^2=4(N-1)$, and this introduces numerical factors into some of 
the previous formulas. For instance, eq.~(\ref{cuarentaidos}) 
becomes\footnote{In order to have the same 
Haggedorn temperature for open and closed strings we keep $\a'=1/2$.} 
\beq
n=N-N'=\frac{1}{4}\big(M\,k^0-\half m^2\big)\ .
\eeq

The computations proceed along the same lines and we will merely 
quote the main results. For $m\ll M$ the energy distribution of 
emitted strings is given by 
\beq\label{cincuentaidos}
d\Gamma_{c}(k^0)\approx \frac{1}{64}A_{D} 
\,g_{c}^2M\,k^0 e^{-k^0/T_{H}}\,k^{D-2}d k\ .
\eeq
Besides the trivial numerical factor, there are two important 
differences with the analogous formula~(\ref{treintaisiete}) for open strings. 
These are the presence of the mass $M$  and the factor $k^{0}$. 
The latter can be seen as a grey body correction to black body 
spectrum, similar to the ones found in other contexts~\cite{gbf}. The factor of 
$M$ indicates that, unlike open strings, closed strings radiate 
proportionally to their mass.

Even more striking is the difference regarding the \emph{total} 
emission rate of strings of mass~$m$. Instead of~(\ref{cuarentaiuna}) we find
\beq\label{cuarentaitres}
\frac{\d\Gamma_{c}(m)}{\d m}\approx  
A''_{D}\,g_{c}^2M\,m_{R}^{-\frac{D-1}{2}}\ ,
\eeq
where $m_{R}\equiv m(M-m)/M$ and $A''_{D}=2^{-\frac{D+15}{2}}\,\pi^{-\frac{3D-5}{2}}
T_{H}^{-\frac{D-1}{2}}$. The power of $m_{R}$ implies that light 
strings are produced much more copiously than heavy ones. However, the 
factor of~$M$ guarantees that the lifetime is proportional to 
$M^{-1}$, as  for open 
strings.
Similarly, the mean  kinetic energy 
released per decay is still given by~(\ref{treintainueve}).

As in the case of open strings, there is a simple semiclassical 
interpretation for~(\ref{cuarentaitres}), at least in the limit $1\ll m\ll M$. 
In order for the initial 
string to split, one point {\bf 1} has to touch a second point {\bf 2}.
The point {\bf 1} can be anywhere along the length $L$ of the string, and this 
explains  the factor of $M\sim L$. Between {\bf 1} and {\bf 2} there is a 
piece of string of length $l\sim m$, and if we assume that the string 
shape is that of a   random 
walk~\cite{ran,dyv}, the mean distance between the two points will be  
$\sim m^{1/2}$. 
 The probability for the two points to meet is inversely
proportional the corresponding volume $m^{(D-1)/2}$ in $D-1$ space dimensions.
For $m\ll M$, $m_{R}\approx m$, and this explains the mass 
dependence in~(\ref{cuarentaitres}).

\subsection{Decays of light strings}
By ``light strings'' we actually mean those with $M^2=2(N-1)$ not 
large enough to use asymptotic approximations for the decay 
rates.\footnote{For simplcity we will consider only light open 
strings. The generalization to closed strings is trivial.}
Then some of the results obtained for heavy strings may no longer be 
valid. Such is the case with the energy 
distributions~(\ref{treintaisiete}), (\ref{treintaiocho}) and 
(\ref{cincuentaidos}), and with the 
production rates~(\ref{cuarentaiuna}) and (\ref{cuarentaitres}).
However, other formulas do not rely on the asymptotic 
approximation~(\ref{treintaicuatro}) for the mass degeneracies and may still 
be valid for $N$ in the tens or hundreds, depending on the level of precission 
sought. This is  true of eqs.~(\ref{treintaiseis}) and~(\ref{cuarenta}).

 On the other hand, 
for small values of $N$ one can obtain \emph{exact} results by summing 
the whole series in~(\ref{veintisiete}). A simple example which serves to illustrate
the fact, mentioned in Section~2, that $F(N,N')$ counts 
``longitudinal'' states for $D<26$ is the 
decay $2\rightarrow 1+1$ of the symmetric rank-two 
tensor~(\ref{dieciseis}) into two photons.
For photon production by light strings, one simply uses~(\ref{veintisiete})
with $F_{NU}=0$, getting
 
\beq\label{veintinueve}
F(2,1)=\G(1)+\G(0)=(D-2)+1=D-1\ .
\eeq
This case is so simple that  we can  compute the amplitudes by hand and sum the 
probabilities over  all the polarizations of the tensor and remaining  
photon  according to the definition~(\ref{treintaidos}) of $F(N,N')$. We find
\beq
D-1-\frac{26-D}{8(D-1)}\ ,
\eeq
which agrees with (\ref{veintinueve}) only for $D=26$. The reason for 
the discrepancy is the following. At the critical dimension the only 
propagating physical state at mass level two is the rank-two 
symmmetric tensor. However, the ``longitudinal'' state~\cite{gsw}
\beq
|\phi\ra=[\a_{-1}\cdot\a_{-1}+\frac{D-1}{5}p\cdot\a_{-2}+\frac{D+1}{10}(p\cdot\a_{-1})^2]
|p\ra\ \ \ ,\ \ -p^2=m^2=2
\eeq
satisfies the Virasoro conditions and has positive norm for $D<26$
\beq
\la\phi|\phi\ra=\frac{2}{25}(D-1)(26-D)\ .
\eeq
The decay of this state into two photons gives a contribution 
$(26-D)/8(D-1)$ which exactly cancels the discrepancy. This example shows 
that, for $D<26$,  the 
double sum $F(N,N')$ includes states which are not simply obtained by 
truncation of the $D=26$ spectrum.

For emission of states with $m^2>0$ we must 
determine $\I_{\nu}(w)$ by direct expansion of the two point-function 
in powers of $v$ and $w$.
Fortunatelly, we only need to compute a few 
terms in the expansion. The reason is 
that the degeneracies  vanish for negative arguments, and by~(\ref{cincuentaitres}) 
and~(\ref{cincuentaicuatro})  
the non-universal contribution takes the form
\beq
F_{NU}(N,N')=a_{\nu}\,\G\Big[N-\nu( A+1)-\half 
m^2(A^2+A)\Big]+\ldots
\eeq
One can easily show 
that 
\beq\label{veintiocho}
\max\Big[N-\nu( A+1)-\half 
m^2(A^2+A)\Big]=1+\textrm{Int}(\frac{m^2}{8})\ ,
\eeq
so that $\I_{\nu}(w)$ must be known only throught order $O(w^{\nu+1+Int
(m^2/8)})$. One can also  show that  the limit~(\ref{veintiocho}) is approached 
only for some ``soft'' 
decays with very little kinetic energy, i.e. with $M\approx m+M'$, and 
that for most decay modes the non-universal contribution vanishes 
because the lowest power of $w$ in $\I_{\nu}(w)\,w^{\nu A+\half 
m^2(A^2+A)}$ is greater than $N$. 

Since the leading universal term in~(\ref{veintisiete}) is proportional to 
\hbox{$\G(N')=\G(1+M'^2/2)$}, whereas by~(\ref{veintiocho}) the leading 
non-universal term is, at best, proportional to $\G(1+m^2/8)$, non-universal terms 
are important only for soft decays with $m\approx 2 M'\approx 2M/3$ or greater.
Indeed, for $m>2M/3$, $A=0$ in~(\ref{veintisiete}) and 
$F(N,N')=F_{NU}(N,N')$.
This is not surprising, since for $m\geq 2M/3$  the ``emitted string'' is more 
appropriately described as a ``remnant'', and the rate is expected to 
depend on its particular state. 

Most decay modes are accurately described by 
the universal series in~(\ref{veintisiete}) and, since this is a rapidly 
decreasing series, it is often enough to keep the first few terms. 
Moreover, there   are even decay modes for which only the first term in the 
series is non-zero. For $M^2>2\,(m^2+M'^2+1)$ the argument of $\G$ for 
the second term in~(\ref{veintisiete}) is already negative and
the following formula is \emph{exact}
\beq
F(N,N')=(N-N'-\half m^2)\,\G(N')\ .
\eeq

As an illustration of these general remarks we will consider  the production of $m^{2}=2$ states, 
i.e. 
decays of the form $N\to 2+N'$. Here we will work at the critical 
dimension $D=26$. In this case there are non-universal 
contributions and we must first compute $\I_{0}(w)$ and $\I_{1}(w)$ 
from~(\ref{treinta}). The result is 
\beqa\label{treintaiuna}
\I_{0}(w)&=&(1+8w)
( \xi_{\mu\nu}p^{\mu}p^{\nu})^2-16w\,(\xi_{\mu\rho}\xi_{\nu}^{\ \ \rho}p^{\mu}p^{\nu})
+4w\,(\xi_{\mu\nu}\xi^{\mu\nu})+O(w^2)\non\\
\I_{1}(w)&=&-2(w+6w^2)\,
( \xi_{\mu\nu}p^{\mu}p^{\nu})^2+4(w+6w^2)\,(\xi_{\mu\rho}\xi_{\nu}^{\ \ \rho}p^{\mu}p^{\nu})
-4w^2\,(\xi_{\mu\nu}\xi^{\mu\nu})+O(w^3)\ 
\eeqa
where, following the analysis around eq.~(\ref{veintiocho}), we have included terms of 
order $w$ and $w^2$ respectively. The values of the Lorentz 
invariants for different polarizations of the emitted tensor are 
given in Table I, where the components are referred to the emitted 
tensor rest frame,  $1$ is the direction in which the tensor is 
emitted, and $a,b=2,3,\ldots,25$.
\newpage

\centerline{
\begin{tabular}{c|c c c|}
Polarization&$\ (\xi_{\mu\nu}p^{\mu}p^{\nu})^2\ $ & $\ \xi_{\mu\rho}\xi_{\nu}^{\ \ 
\rho}p^{\mu}p^{\nu}\ $ & $\ \xi_{\mu\nu}\xi^{\mu\nu}\ $\\
\hline \hline
$\xi_{ab}$ & $0$ & $0$ & $\frac{1}{2}$\\
$\xi_{1a}$ & $0$ & $\frac{1}{4}x$ & $\frac{1}{2}$\\
$\xi_{11}$ & $\frac{12}{25}x^2$ & $\frac{12}{25}x$ & $\frac{1}{2}$\\
\end{tabular}}
\vskip.5cm
\centerline{{\bf TABLE I.} Values to be used in~(\ref{treintaiuna}) with $x=(1+n)^2/2-M^2$.}

\vskip1cm
\noindent For example, for $7\to 2+3$ we have $n=7-3=4$ 
and~(\ref{veintitres}) gives 
\beq
\I_{4}(w)=3w^4+w^{6}+w^{6}\I_{0}(w)
\eeq
which, using Table I with $x=\half$ yields
\beq\label{cincuentaisiete}
F(7,3)=3\G(3)+\G(1)+\left\{ \begin{array}{lll} 
2\ \ =\ \ 9626 &\ \  \textrm{for} &\xi_{ab}\\ 0\ \ =\ \ 9624 &\ \  \textrm{for} & \xi_{a1}\\
2\ \ =\ \ 9626 &\ \  \textrm{for} & \xi_{11}\ \ .
\end{array} \right. 
\eeq
The other possible non-tachyonic decay modes of the form $7\to 2+N'$ are $7\to 2+2$ 
and $7\to 2+1$. The corresponding sums are universal and given by
\beqa
\I_{5}(w)&=&4w^5+O(w^8)\ \ \ ,\ \ \     F(7,2)=4\,\G(2)=1296\non\\
\ \I_{6}(w)&=&5w^6+O(w^{10})\ \  ,\ \ \     F(7,1)=5\,\G(1)=120\ \ \ \ .
\eeqa
Note that out of the three decay modes only $7\to 2+3$, which is the 
one with the smallest kinetic energy, has non-universal contributions 
and that these are very small, of the order of $0.02\%$. A similar 
computation for $3\to 2+1$ and $4\to 2+1$ yields non-universal 
contributions of $8\%$ and $4\%$ respectively. This confirms that 
these contributions are significant only when the emitted particle 
carries a large fraction of the original mass. Also note that the 
relative contribution of the second term in~(\ref{cincuentaisiete}) is rather 
small, of the order of $0.25\%$. Similar examples show 
that~(\ref{cuarenta}), which uses only the first term in the 
solution, 
gives very accurate results for $N$ as low as $10$ or $20$.
 
\newpage

\section{Discussion}

In this paper we have considered the problem of computing the 
\emph{averaged semi-inclusive} rates for emission of  
\emph{specific} string states of mass $m$. This requires the 
evaluation of the double sum
\beq
F(N,N')= \sum_{ \Phi_i|_{N}}\sum_{ \Phi_f|_{N'}} \big|
 \langle \Phi_f |  V(k)|\Phi_i \rangle \big|^2\ ,
\eeq 
where $V(k)$ is the vertex operator for a string state of mass $m$.
Our solution   involves a recursion relation~(\ref{veintidos}) 
which is a linear difference equation and depends only on the mass $m$ 
of the emitted string, with the  details of the specific emitted state 
entering only as ``initial conditions''. The universal 
series in the solution~(\ref{veintitres})
\beq
\I_{n}(w)=\sum_{p=1}^A \Big(n-m^2(p-\half)\Big)w^{np-\half 
m^2(p^2-p)}+\ldots
\eeq
suggests some sort of ``deformation'' of Plank's distribution for 
massless strings
\beq
\I_{n}(w)=\frac{n w^n}{1-w^n}=\sum_{p=1}^\infty n\, w^{np}\ .
\eeq
It would be interesting to check if this series is truly universal and appears in other 
averaged semi-inclusive quantities such as $N$-point functions, or is 
just particular to the $3$-point functions considered here.

Being able to compute averaged semi-inclusive $4$-point functions would also be interesting 
in its own right. They would give us the ``form factors'' of 
\emph{typical} fundamental strings, with direct information on their 
shape and size. This would be an alternative approach, along the lines 
of~\cite{bo}, to the one 
pursued in~\cite{dyv} for the determination of the sizes of 
fundamental strings~\cite{sus}. One could also try to compare the form 
factors with the scattering properties of black holes, and see it 
some of them can be adscribed to a ``decoherence'' process~\cite{coh}.

Our analysis has shown that the decoherence procedure of 
averarging over all initial states of mass $M$ is very efficient in 
suppressing the dependence on the state of the emitted string, 
unless this string carries a very substantial fraction of the 
original mass. In fact, most decay modes are well described by the 
first term in the universal series. Dividing the approximate 
expression~(\ref{treintaiseis}) for $F(N,N')$ by $\G(N)\G(N')$ yields the 
``averaged interaction rate''
\beq
<\big|\langle \Phi_f |  V(k)|\Phi_i \rangle \big|^2\!>\approx 
 (N-N'-\half m^2)\frac{1}{\G(N)}=\frac{k\cdot p'}{\G(N)}
\eeq
for the reaction $M\to m+M'$ with $p=k+p'$. Note that in this case we 
have  averaged over the \emph{three} strings involved. For closed strings the 
result is similar
\beq
<\big|\langle \Phi_f |  V(k)|\Phi_i \rangle \big|^2\!>\approx 
\frac{(k\cdot p')^2}{\G(N)^2}
\eeq
and was obtained in~\cite{lys} by a different method. These results 
are very simple, and suggest that similar simplifications might  take 
place for higher order interactions.

Our results for the production rates of strings of mass $m$ by heavy 
open and closed strings
\beq\label{cincuentaiseis}
\frac{\d\Gamma_{o}(m)}{\d 
m}=A'_{D}g_{o}^2  \ \ ,\ \  
\frac{\d\Gamma_{c}(m)}{\d m}= 
A''_{D}\,g_{c}^2M\,m_{R}^{-\frac{D-1}{2}}\ \  ,\ \ 
m_{R}\equiv\frac{m(M-m)}{M}
\eeq
have very simple semiclassical interpretations and provide a microscopic 
basis for  the type of heuristic interactions used in the Boltzmann 
equation approach to hot string gases~\cite{hss,dbr}. In particular, 
the small value of the mean kinetic energy obtained in~(\ref{treintainueve}) gives 
support to the assumption that the total length of the strings is approximately
conserved.  It is interesting that the mass dependences 
in~(\ref{cincuentaiseis}), which from a semiclassical viewpoint are a 
consequence of the random walk structure of long strings, arise 
microscopically as a result of the interplay between  mass 
degeneracies and  phase space factors. Their effects cancel out for 
open strings, giving rise to a flat spectrum. The decay rates for typical 
open
strings computed in this paper agree, 
within a factor of $2$, with those computed for strings in very 
special states~\cite{cdr,dyp}.

For open strings on a Dp-brane the spectrum is no longer flat. The string 
can oscillate in $D$ dimensions, but its endpoints are bound to a 
$p+1$ manifold. The phase space factor in~(\ref{cuarenta}) 
becomes $k^{p-2}$, and instead of a flat spectrum we find a rate 
proportional to $m_{R}^{-d/2}$, where  $d\equiv D-p-1$ is the number 
of ``extra dimensions'' orthogonal to the brane. This may be relevant 
to the decays of ``string balls'' in brane-world scenarios~\cite{rob}.
It would be interesting to study the decay properties of strings in 
other situations involving D-branes, such as those considered 
in~\cite{bar,dbr}. Another possible extensions along these lines 
would be to include the effect of 
external fields and non-conmutative limits of the theory~\cite{bac} on 
the decay rates.

\vskip2cm
\centerline{\bf ACKNOWLEDGMENTS}
\vskip.5cm
It is a pleasure to thank J.M.~Aguirregabiria, I.L.~Egusquiza, R.~Emparan,  
Jaume Gomis, Joaquim Gomis,  M.A.~Valle-Basagoiti and M.A.~V\'azquez-Mozo for useful and
interesting discussions. This work  has been  supported in part by
the Spanish Science Ministry under Grant AEN99-0315 and by a University of 
the Basque Country Grant UPV-EHU-063.310-EB187/98.

\appendix
\section{}
Here we will compute the integral~(\ref{cuarentaicuatro}) which appears 
in the derivation of the recursion 
relation. 
The scalar two-point 
function on the cylinder~$\psi(v,w)$  is related to the Jacobi 
 $\vartheta_{1}$ function by\cite{gsw}
\beq\label{auno}
\psi(v,w)=\frac{2\pi 
i}{\tt}\,\frac{\vartheta_{1}(\nu|\tt)}{\vartheta'_{1}(0|\tt)}\ \ \ ,\ \ 
\tt\equiv-\frac{2\pi i}{\ln w},\ \ \ ,\ \ \nu\equiv\frac{\ln v}{\ln w} 
\eeq
and using the product formula  for the theta function  yields the 
following  expression  
\beq\label{acinco}
\psi(v,w)=-\frac{2\pi}{\ln q}\sin \pi \nu\prod_{1}^\infty\frac{1-2q^{2n}\cos 
2\pi\nu+q^{4n}}{(1-q^{2n})^2}\ ,
\eeq
where $q=e^{i\pi\tt}$. The function $\psi(v,w)$ has zeroes for $v=w^n$ with $n=0,\pm 
1,\pm2, \ldots$, and expanding~(\ref {acinco}) around $v=1$ gives 
\beq
\psi(v,w)=-\ln (v)+O((\ln v)^3)=-\tau+O(\tau^3)\ ,
\eeq
where  $\tau$ is the euclidean proper time used throughout the 
paper and should not be confused with the modular parameter $\tt$ 
in~(\ref{auno}). This fixes the behavior of the function 
$\hat\psi(v,w)$ near  $v=1$
\beq
\hat\psi(v,w)=\sqrt{v}\exp\left( {-\ln^2 v\over2 \ln w}\right)\, \psi(v,w)=
-\tau(1+\half\tau)+O(\tau^3)
\eeq
and implies
\beqa\label{ados}
\la e^{-ik.X_{o}(1)} e^{ik.X_{o}(v)}\ra&=&\hat\psi(v,w)^{m^2}=
\tau^{m^2}\Big(1+\frac{m^2}{2}\tau+O(\tau^2)\Big)\non\\
\la \del_{\tau} \Xm_{o}(1) \del_{\tau}\Xn_{o}(v) 
\ra&=&\eta^{\mu\nu}\del^2_{\tau} 
\ln \hat\psi(v,w)=- \Om(v,w)=-\frac{1}{\tau^2}+O(1)\ .
\eeqa
Each additional proper time derivative contributes a negative power of $\tau$, 
and (\ref{ados}) together with the mass shell condition implies the 
following form for the correlator 
\beq\label{atres}
\la {V'}^{\dagger} (k,1) \ V'(k,v) \ra 
=\frac{A}{\tau^2}(1+\frac{m^2}{2}\tau)+O(1)\ .
\eeq

The value of $A$ can be found by noting that computing the leading 
singularity of the correlator with the help of~(\ref{ados}) is 
equivalent to finding the leading term in the OPE of ordinary 
(unprimed) vertex 
operators. But this term is given by 
\beq\label{acuatro}
V_{i}^{\dagger}(0)V_{j}(\tau)=\frac{\la\la\bar{i}|j\ra}{\tau^2}+\ldots
\eeq
where $\la\la\bar{i}|j\ra$ is essentially Zamolodchikov's inner 
product, which is related to the familiar quantum 
mechanical inner product\footnote{See chapter 6 of~\cite{pol} for  
details.} by   
$\la\la\bar{i}|j\ra=\la{i}|j\ra$. Comparing (\ref{atres}) and 
(\ref{acuatro})  implies $A=1$ for normalized states.
Computing the integral  gives 
\beq
\oint_{C}\frac{dv}{v}v^n\la {V'}^{\dagger} (k,1) \ V'(k,v) 
\ra=\oint_{0}d\tau 
e^{n\tau}\frac{1}{\tau^2}(1+\frac{m^2}{2}\tau)=n+\frac{m^2}{2}\ .
\eeq

\section{}
In this Appendix we present an approximate expression for the mass level 
degeneracy which is  more accurate than the usual asymptotic 
formula~(\ref{treintaicuatro}). We have used it in our numerical computations in 
Section~5, and   find it  particularly useful for the range of values of $N$ 
where a direct power expansion of the partition function is not
practical, while the asymptotic formula is not yet accurate enough.

The basic idea is to  compute the first corrections to the 
saddle-point aproximation for 
\beq
\G(N,D)=\oint\frac{d w}{w} w^{-N} f(w)^{2-D}\ \ \ ,\ \ \ 
f(w)=\prod_{n=1}^\infty(1-w^n)\ .
\eeq
Since the saddle-point  is close to $w=1$, one has to   use the modular 
transformation
\beq
f(w)=\bigg(\frac{-2\pi}{\ln w}\bigg)^{1/2} w^{-1/24} q^{1/12}f(q^2)\ \ 
\ ,\ \ \ q\equiv e^{2\pi^2/\ln w}
\eeq
which gives
\beq
\ln f(w)=\half\ln 2\pi-\half\ln\b+\frac{1}{24}\b-\frac{\pi^2}{6\b}+O
\big(e^{-4\pi^2/\b})
\eeq
where we have  made the change of  variable $w=e^{-\beta}$. Keeping all
corrections of order $O(N^{-1/2})$ yields the following result
\beq\label{buno}
\G(N,D)\sim\Big(\frac{24}{D-2}\Big)^{1/2}\frac{e^{a(x)}}{(2\pi 
b(x))^{1/2}}\Bigg(1+\frac{c(x)}{D-2}\Bigg)\ ,
\eeq
%
%
where
\beqa
a(x)&\equiv&(N-\frac{D-2}{24})\frac{1}{x}+\frac{D-2}{24}\Big(4\pi^2 
x-12\ln x-12\ln 2\pi\Big)\non\\
b(x)&\equiv&4x^2(2\pi^2x-3)\non\\
c(x)&\equiv&\frac{9}{2}\frac{4\pi^2x-3}{(2\pi^2x-3)^2}-\frac{45(\pi^2x-1)^2}{(2\pi^2x-3)^3}
\eeqa
and the position of the saddle-point  is given by
\beq
x\equiv\frac{1}{\b_{0}}=\frac{1}{2\pi^2}\Bigg\{3+\sqrt{9+\pi^2\Big(\frac{24N}{D-2}-1\Big)}\Bigg\}\ .
\eeq

The following table gives a few sample values for the degeneracies, 
and we can compare the exact results for $\G(N,D)$ with those obtained 
from~(\ref{buno}). Note that the agreement is very good for $N$ as 
low as $20$, and  even the results for $N=1$ are acceptable.

\vskip1cm

\centerline{
\begin{tabular}{|c||c|c|c|c|}
\hline
$\ \ \textrm{N}\ \ $ & $\G(N,26)$ & (\ref{buno}) & $\G(N,16)$ & (\ref{buno})\\
\hline \hline
$1$ & $24$ & $23.819$ & $14$ & $13.949$ \\
$5$ & $176,256$ & $176,271$ & $19,754$ & $19,759$ \\
$20$ & $\ 2.16109\cdot 10^{14}\ $  & $\ 2.16118\cdot 10^{14}\ $ & 
$\ 3.41780\cdot10^{11}\ $  & $\ 3.41806\cdot10^{11}\ $ \\
$100$ & $2.43478\cdot 10^{40}$ & $2.43480\cdot 
10^{40}$ & $7.90265\cdot10^{31}$ & $7.90276\cdot10^{31}$ \\
\hline
\end{tabular}}
\vskip.5cm
\centerline{{\bf TABLE II.} Sample values of $\G(N,D)$}

\vskip.7cm
This should be compared with the usual asymptotic approximation
\beq
\G(M)\sim(2\pi 
T_{H})^{-\frac{D-1}{2}}M^{-\frac{D+1}{2}}e^{M/T_{H}}\ \ ,\ \ \
T_{H}=\frac{1}{\pi}\,\sqrt{\frac{3}{D-2}}
\eeq
which,  for $N=100$, gives $4.79\cdot10^{40}$ for $D=26$ and 
$9.20\cdot10^{31}$ for $D=16$.

\end{document}